\RequirePackage{fix-cm}
\documentclass[twocolumn]{svjour3}          
\smartqed  

\usepackage{cite}
\usepackage{amsmath,amssymb,amsfonts}
\usepackage{algorithmic}
\usepackage{graphicx}
\usepackage{textcomp}
\usepackage{xcolor}
\usepackage{graphicx}
\usepackage{subcaption}
\usepackage{csquotes}
\begin{document}

\title{Photonic Network Coding and Partial Protection for Optical Core Networks: Two for a Tango 
}


\author{Dao Thanh Hai   
}


\institute{F. Author \at
	Post and Telecommunication Institute of Technology \\
	\email{haidt102@gmail.com} 
}

\date{Received: date / Accepted: date}

\maketitle

\begin{abstract}
The age of acceleration is taking place, driven by the revolutionary digital transformation creating basically a digital version of our physical world and the currency in that digital space is data. Massive amount of data has been generated ranging from wearable devices monitoring our physical health every single millisecond to autonomous vehicles generating roughly 5Tb hourly to even astronomical activities producing an order of Exabytes on daily basis and then ultra-broadband Internet comes into play, moving such data to the cloud. Internet traffic therefore has been experiencing explosive growth and in this context, optical transport networks forming the backbone of the Internet are pushed for transformation in system capacity. While the intuitive solution of deploying multiple fibers can address the pressing demand for increased capacity, doing so does not bring improvement in economic of scales in terms of cost, power consumption and spectral efficiency. This necessitates for a different approach so that the fiber capacity could be utilized in a more efficient manner. In this paper, we focus on innovative techniques, that is, photonic network coding and partial protection, to reduce the effective traffic load in order to achieve greater capacity efficiency for optical transport networks. Specifically, the application of network coding is examined by upgrading the functionalities of intermediate nodes with all-optical processing (i.e., encoding and decoding) capabilities. Besides, partial protection relying on the premise of providing just enough bandwidth in case of failure events is investigated for saving the redundant protection capacity. That it takes two to tango, combining photonic network coding and partial protection therefore bring to light new opportunities and challenges. In mining such new avenue, we present insights on how to derive compounding gains to maximize spectral efficiency via a case study. 

\end{abstract}

\section{Introduction}
\label{intro}
Data is here, data is there and data is all around you. Gone are the days when data was on the scale of Megabytes and Gigabytes and today it is the era of Terabytes and Petabytes. Indeed, thanks to the proliferation of super-cheap computer chips and the ubiquitous wireless networks, a revolutionary process is taking place, turning anything, from something as tiny as a pill to something as sizable as an aeroplane, into a part of the Internet of Things \cite{new3}. Such massive amount of data generated and circulated in Internet have given the rise to explosive growth of Internet traffic. According to a recent report from Cisco, Global Internet traffic will grow 3.2-fold from 2016 to 2021 with a compound annual growth rate of $26\%$ and the situation becomes more severe for the busy hour Internet traffic whose growth will be 4.6-fold from 2016 to 2021 \cite{Cisco20}. In this context, optical transport networks forming the backbone of the Internet have been pushed for radical transformation in system capacity to accommodate insatiable traffic demands in a sustainable way \cite{ir4}. \\

For many years, the capacity of an optical fiber has been viewed as being almost infinite and indeed, the single channel capacity has undergone leap-and-bound growth with a spectacular rise from 2.5 Gb/s in around 1990 to beyond 1 Tb/s in 2020 \cite{20years, new1}. Such factor of 400-fold increase in a span of roughly 30 years is thanks to convergence of remarkable advances in electronic, photonic and digital signal processing technologies \cite{hai_iet, hai_wiley, hai_nics, EON}. In accompany with system capacity expansion, it has to be noted that there has been more than 3 orders of magnitude decrease in cost per $Gb/sec \times km$, from roughly $1,000$ $\textdollar /(Gb/sec \times km)$ down to less than  $0.50$ $\textdollar /(Gb/sec \times km)$ \cite{20years, new4}. However fiber throughputs are now coming close to the hard limitation determined by the well-established nonlinear Shannon limit, so network operators are exploring innovative ways to enlarge the number of parallel optical paths by means of spatial-division multiplexing (SDM) \cite{new2}. There have been some technological options for implementing SDM including lighting up more fiber-pairs per link instead of single fiber-pair, or alternatively making use of multi-core and/or few-mode fibers. While the first option is intuitive, it does not result in any improvement on the cost and energy per bit/sec. The second option is expected to boost the system capacity more than an order of magnitude and yet a lot of challenges in implementation and tremendous capital costs remain to be addressed. As an economical consequence, such option is unlikely to be deployed in the near-term. Different from technological approaches aiming at expanding the system capacity with major investments, architectural strategies do not rely on the system capacity increase to handle more traffic but rather, reduce the effective traffic load so that more traffic can be carried \cite{sla1, qos1}. \\

Network Coding (NC) and Partial Protection (PP) have been emerging as efficient architectural strategies to address the explosive traffic growth in the near-term scale by means of better utilizing the fiber capacity \cite{hai_comletter, hai_oft, hai_access, hai_comcom, hai_comcom2, hai_systems, nc_others1, nc_others2, nc_others3}. Network Coding, originally proposed in \cite{NC}, has soon become a radical technique in networking to achieve higher throughput, security and capacity. The main idea is to reduce the traffic load in network by encoding signals at favorable conditions and by transmitting such encoded signal rather sending individual ones, less capacity is required \cite{hai_rtuwo, hai_springer, hai_springer2}. Partial protection, on the other hand, has the potential of saving spectrum resources by offering just enough protection capacity for demands instead of over-provisioning as in the conventional approach \cite{sla2, qos2}. In this paper, we present, for the first time, original perspectives on how Network Coding and Partial Protection could generate remarkable spectral savings in optical networks realm. More interesting is brought in the case when Network Coding meets Partial Protection and how to optimize the such additive impact. Our analysis on this unique situation could serve as a guideline to network operators to realize extra benefits from combining Network Coding and Partial Protection. \\

The structure of the paper is organized as followed. In Sect. 2, related works are briefly presented on the emerging of network coding and QoS-aware protection for optical networks. Next, in Sect. 3 and Sect. 4, we showcase the niche for leveraging photonic network coding and partial protection to achieve greater capacity efficiency, respectively. Section 5 is dedicated to an innovate case study where photonic network coding meets partial protection and we provide a detailed analysis on how to derive compounding gain in such unique cases. \\

\section{Related Works}  
Network coding (NC) has been a paradigm shift in communication networks by bringing the capability of in-network computing. The central idea is that intermediate nodes, instead of simply storing and forwarding data as adopted in traditional networking paradigm, is empowered to manipulate transiting data and then forward such (non-) linearly combined data to its output. The successes of network coding has been remarkable and therefore has been a \textit{de facto} in future wireless networks \cite{s7}. In fact, early attempts of exploiting NC for optical networks have been proposed in \cite{s1, s2} for the scenario of protection where multiple signals could be favorably encoded. Other than protection realm, NC have been particularly well-suited for the multi-cast transmission and this was the focus of the work in \cite{s5, s6}. It has to be noted that the majority of existing works on NC for optical networks have been focused on performing NC in electrical domain in combination with optical-electrical-optical network architecture and this was due to the immature of photonic signal processing in the past. In recent years, though, the rapid advances in all-optical signal processing have renewed the interest of applying NC in optical networks for achieving greater capacity efficiency with a special focus on performing NC in photonic domain. The work in \cite{s3} has taken advantage of NC in optical transport networks for provisioning cloud radio access network services with greater capacity efficiency while maintaining fast responses. The use of photonic NC has also been extended to security services at physical layer with pioneering works in \cite{s4}. Applications of NC have also been found in mm-wave radio-over-fiber networks \cite{nc_others1}, in visible light communications \cite{nc_others2, nc_others3}, and in passive optical networks (PON) \cite{nc_others4}. Data-center networking has been an active area for applying NC to reduce traffic \cite{nc_others5, nc_others6, nc_others8}. \\

Partial protection has been driven by the observation that roughly $30\%$ traffic traveling in optical fiber belongs to premium class while $70\%$ remaining are best-effort type and hence can be temporarily discarded on occasion of failures \cite{sla1, sla2, sla3}. This means that instead of providing full protection, just some part of the traffic could be delivered to receivers upon failure events and by reducing the amount of protection requirement, less spare resources might be needed, permitting saving of both operational and capital expenditures. Partial protection had been raised in the context of traditional wavelength-division multiplexing (WDM) networks with pioneering works from \cite{pp1, pp2, pp3}. However, due to the fixed nature of WDM networks, there has been little room for further improvement with partial protection and consequently the concept of partial protection had been somehow faded. The development of elastic optical networks (EONs) with adaptive resource allocation according to specific traffic requirement and/or transmission qualities opens up therefore new opportunities for a re-consideration of partial protection In fact, a closely related strategy is called, quality-of-service provisioning, in EONs have been studied in \cite{pp5, pp6, pp7, pp8}. Recent works in \cite{qos1, qos2} have been pushing forward the applicability of partial protection in EONs by presenting an optimal mathematical framework to maximize partial protection benefits. \\

It is important to note that both network coding and partial protection techniques belong to the architectural approaches to address the continued explosive traffic in optical core networks. It means that rather than expanding system capacity, such architectural approaches aim at reducing effective traffic load in the network so as to save spectral capacity for future traffic. To the best of our survey, existing works in the literature have so far addressed the use of each technique separately and this necessitates for a question of whether partial protection could be jointly applied with network coding to achieve additive impact. We aim to address this question in this paper. \\ 

Our original contribution is indeed a proposal for combining photonic network coding and partial protection in elastic optical core networks based on flexible channel spacing and how such combination can result in greater capacity efficiency than making use of individual technique. Furthermore, we pinpoint an interesting issue on how to maximize the combination impact, that is, the selection of protection level for each demand so that on one hand, the overall partial protection specification is met and on the other hand, the coding opportunity is optimized.

\section{Network Coding for Greater Capacity Efficiency}

In this part, we showcase a niche of applying NC in optical transport networks that could bring about major spectrum savings by reducing the effective traffic load in the network. Let's consider an elastic optical network shown in Fig. 1 and assume that there are two traffic demands with different bit-rates, from node $A$ to node $Z$ and from node $B$ to node $Z$ with 100 Gbps and 50 Gbps respectively. We also assume the modulation format to be used in entire network is 16-QAM with polarization multiplexing and the standard spectrum slice width of 6.25 GHz. Having such assumptions, provisioning those demands with dedicated protection then involves finding a pair of link-disjointed route and allocate spectrum slices along such routes for each demand as illustrated in Fig. 1. It can be seen that due to the non-overlap spectrum constraint, three spectrum slices are needed along the link $XE$ and $EZ$. This is the traditional approach without network coding operations. \\

\begin{figure}[!ht]
	\centering
	\includegraphics[width=\linewidth, height = 5.5cm]{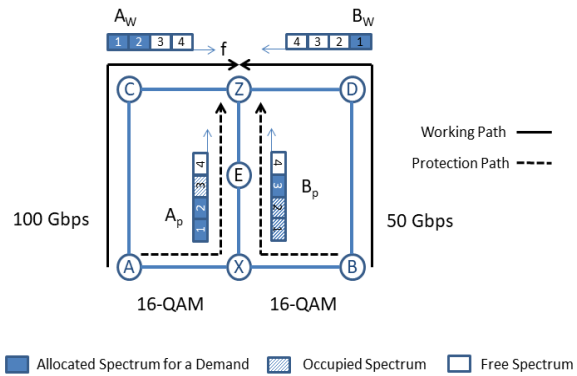}
	\caption{Traditional Approach without Network Coding Functions}
	\label{fig:i1}
\end{figure}

Now let's turn to the situation when node $X$ is armed with encoding capability. In particular, supposing that at node $X$, two protection traffic $A_p$ and $B_p$ could be XOR-encoded and the encoded signal $A_p \oplus B_p$ operates at 100 Gbps, occupies two spectrum slots and goes all the way to the receiving node $Z$. It is noticed that by doing so one spectrum slice is saved along the route $XE$ and $EZ$, making up roughly $30\%$ spectral savings. At the destination node $Z$, three signals are received including $A$, $B$ and $A_p \oplus B_p$ and if one signal is lost due to a failure, it can be recovered (almost) immediately from the other signals that are received. The encoding and decoding process is clearly shown in Fig. 2. The underlying idea behind this trick is that by combining signals at specific nodes, rather than duplicating the signals end-to-end, network resources may be used more efficiently.

\begin{figure}[!ht]
	\centering
	\includegraphics[width=\linewidth, height = 5cm]{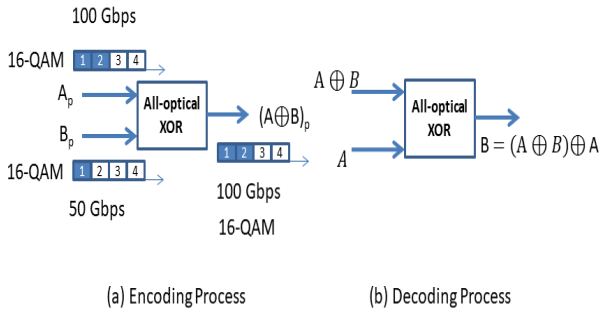}
	\caption{Network Coding-enabled Approach}
	\label{fig:i2}
\end{figure}

\section{Partial Protection for Greater Capacity Efficiency}
Partial protection is inspired by the observation that the premium traffic accounts for roughly $30\%$ of the whole traffic traveling in optical fiber links today whereas the remaining is best-effort. While the premium type demands rapid recovery upon failures, the best-effort one could be dropped and restored after a few tens of seconds without incurring any penalties. Nevertheless, conventional approach treats both traffic types with the same level of 1+1 expensive fast protection and this opens up new opportunities to offer protection bandwidth in a more granular manner via a concept called partial protection. Besides, partial protection is also backed by the case that a service can accept some degradation (i.e., decreased rate) upon a failure in exchange for a reduced cost and network operators, for the sake of optimizing network resources and reduce cost, may wish to support degraded services. \\

A case for illustration of applying partial protection is examined in this part. In the traditional full protection shown in Fig. 3, four spectrum slices are needed along the protection route of that 100 Gbps demand from node A to node B. However, if we take into account the fact that such 100 Gbps traffic consists of two types, namely, premium and best-effort one, new opportunities for spectrum saving arise. For example, in Fig. 4, assuming that there is only half traffic of premium type and thus, needs to guarantee rapid recovery upon failures while the rest can be discarded temporarily. Doing so therefore saves the protection bandwidth of two spectrum slices along the protection route, offering a remarkable $50\%$ spectral savings. Clearly the amount of saving is dependent to the amount of protected traffic and the less protected traffic is, the more saving it could be achieved.

\begin{figure}[!ht]
	\centering
	\includegraphics[width=\linewidth]{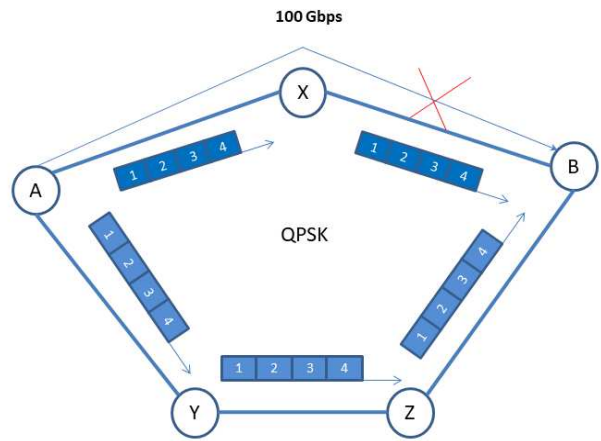}
	\caption{Traditional Full Protection}
	\label{fig:i3}
\end{figure}

\begin{figure}[!ht]
	\centering
	\includegraphics[width=\linewidth]{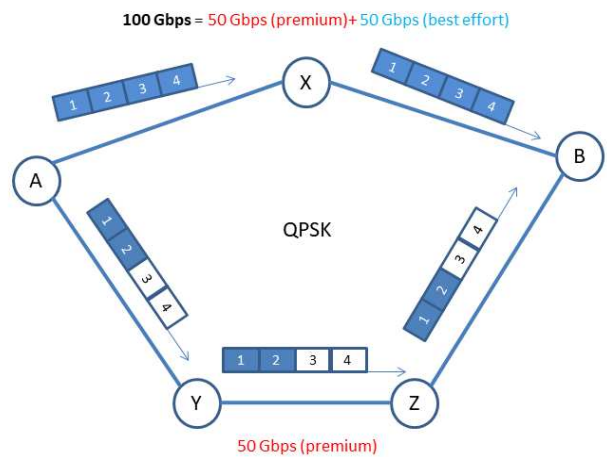}
	\caption{Partial Protection Approach}
	\label{fig:i4}
\end{figure}

\section{When Network Coding meets Partial Protection: An Interesting Game Arises}
As highlighted in above sections about the capability of network coding and partial protection in reducing the effective traffic load in the network and therefore utilizing better fiber capacity, the next step pinpoints to an interesting situation if both network coding and partial protection is applied. In such cases, we are interested in a guideline so that the optimal spectral gain could be achieved and it will be tackled in this part.  \\

Consider the scenario as shown in Fig. 5  for an elastic optical network adopting the modulation format QPSK (polarization multiplexing) for the whole network and a standard spectrum slice width of 6.25 GHz. Assuming that there are two traffic demands from node $A$ and node $B$ to node $Z$ with 100 Gbps and 150 Gbps respectively. With such assumption, four spectrum slices are occupied by the first demand while the second demand requires six slices (i.e., each spectrum slice carries 25 Gbps). For this scenario, in addition to the opportunities of reducing protection traffic by means of partial protection, the fact that $A_p$ and $B_p$ shares the same link $XZ$ paves the way also for exploiting network coding. Table I draws a performance comparison measured by number of spectrum slices on link $XZ$ if network coding and partial protection is separately applied. It can be observed that applying network coding alone brings about $40\%$ spectral savings while for applying partial protection, the saving is dependent on the amount of protected traffic. For a total of 250 Gbps traffic occupying 10 spectrum slices, each $10\%$ reduction in protection traffic is translated to one spectrum slice saving and it is detailed in Table 1.  \\

\begin{figure}[!ht]
	\centering
	\includegraphics[width=\linewidth]{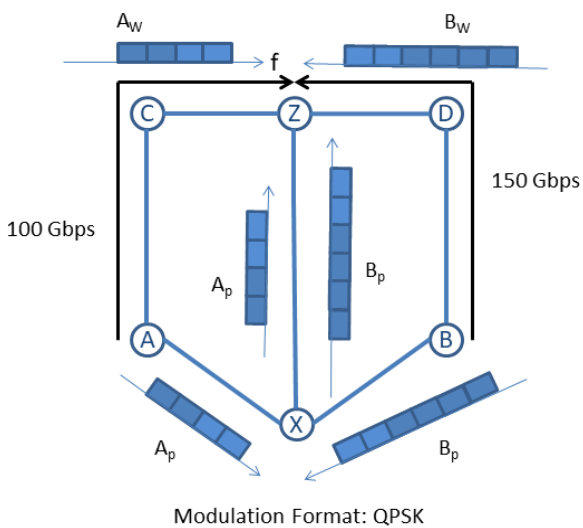}
	\caption{An Exemplary Case to Combine Network Coding and Partial Protection}
	\label{fig:i5}
\end{figure}

\begin{table}[ht]
	\caption{Performance Comparison when Applying Network Coding and Partial Protection Separately}
	\label{tab: r1}
	\centering
	\begin{tabular}{|c|c|}
		\hline
		\scriptsize{Scenario} & \scriptsize{$N^o$ spectrum slices on link $XZ$} \\
		\hline \hline
		
		\scriptsize{Full protection w/o NC} & 10 \\
		\scriptsize{Full protection with NC} & 6 \\
		\hline \hline 
		\scriptsize{Partial Protection ($90\%$) w/o NC} & 9 \\
		\scriptsize{Partial Protection ($80\%$) w/o NC} & 8 \\
		\multicolumn{2}{|c|}{...} \\
		\scriptsize{Partial Protection ($10\%$) w/o NC} & 1 \\
		\hline 
	\end{tabular}
\end{table}

Let us now focus on the joint application of network coding and partial protection. Of the total 250 Gbps traffic by two demands, if partial protection is applied and the choice then arises on how much reduction is applied on the first demand (100 Gbps) and how much the reduction is applied to the second demand (150 Gbps) so as to optimally profit from network coding operation. For convenience, we assume the reduction in traffic occurs at a step of 25 Gbps (i.e., corresponds to one spectrum slice and $10\%$ total traffic). \\

Basically, the process determining the optimal protection traffic for each demand consists of two steps. The first step is to list all possible configurations of providing protection level for each demand so that the overall protection traffic is met. Next, for each configuration, the second step is to perform network coding for protected traffic of each demand and determine the amount of needed spectrum slices. After performing encoding for all possible configurations, an optimal one is identified in terms of minimizing number of used spectrum slices. \\

For illustration, consider the $10\%$ degradation case ($90\%$ protected), we have two options, that is, either, that $10\%$ is reduced on first demand ($A_p$) or on the second demand ($B_p$). Since the second demand carries higher bit-rate and consequently occupies larger amount of slices than the first demand, the outcome signal of encoding second demand and first demand will occupy the same amount of slices as with the second demand. This observation suggests that if $10\%$ is reduced by the second demand, the number of spectrum slices needed on link XZ after encoding will be five. Otherwise, if that $10\%$ is reduced by the first demand, the number of spectrum slices needed on link XZ after encoding will be still six. By same way of reasoning, Table 2 provides the optimal configuration with respect to relative amount of protection traffic and the corresponding number of slices on link XZ after performing network coding. 

\begin{table}[ht]
	\caption{Performance Comparison when Network Coding is Optimally Combined with Partial Protection}
	\label{tab: r2}
	\centering
	\begin{tabular}{|c|c|c|}
		\hline
		Scenario & Configuration & $N^o$ slices\\
		\hline \hline
		$90\%$  & 100 + 125 & 5 \\
		$80\%$  & 100 + 100 & 4 \\
		$70\%$  & 100 + 75 (75 + 100) & 4 \\ 
		$60\%$  & 75 + 75 & 3 \\ 
		$50\%$  & 75 + 50 (50 + 75) & 3 \\ 
		$40\%$  & 50 + 50  & 2 \\
		$30\%$  & 50 + 25 (25 + 50)  & 2 \\
		$20\%$  & 25 + 25  & 1 \\
		$10\%$  & 25 + 0 (0 + 25)  & 1 \\
		\hline 
	\end{tabular}
\end{table}

Such optimal configuration provided in Table 2 allows network operators to decide the best amount of reduction for each demand so as to achieve the maximal benefit of network coding operations. \\

\section{Summary}
Inspired by the observation that optical transport networks have been under critical pressures due to the explosive traffic growth in \enquote{data, data, data} era, we have proposed architectural strategies including Network Coding and Partial Protection to better utilize the fiber capacity and hence, achieve greater capacity efficiency. This is possible thanks to the unique opportunity enabled by Network Coding and Partial Protection to reduce the effective traffic load in the network. We have pinpointed and analyzed the original prospect when Network Coding meets Partial Protection and how to optimize such additive impact. That analysis might play a guideline role for network operators in operating infrastructure in a more spectrally efficient manner. \\

It has, nevertheless, be noted that our proposed solution does not aim to replace long-term technological innovations of expanding system capacity, but rather serve as consolatory and complementary solutions to mitigate the approaching of capacity crunch. Our proposal should therefore be paid attention to before jumping to next curve of technological paradigm such as spatial-division multiplexing framework.    


%
\section*{Conflict of interest}
The authors declare that they have no conflict of interest.


\bibliographystyle{spmpsci_modified}      
\bibliography{revised}   

\begin{thebibliography}{10}
\providecommand{\url}[1]{{#1}}
\providecommand{\urlprefix}{URL }
\expandafter\ifx\csname urlstyle\endcsname\relax
  \providecommand{\doi}[1]{DOI~\discretionary{}{}{}#1}\else
  \providecommand{\doi}{DOI~\discretionary{}{}{}\begingroup
  \urlstyle{rm}\Url}\fi

\bibitem{new3}
Babarczi, P., Tapolcai, J., Pašić, A., Rónyai, L., Bérczi-Kovács, E.R.,
  Médard, M.: Diversity coding in two-connected networks.
\newblock IEEE/ACM Transactions on Networking \textbf{25}(4), 2308--2319
  (2017).
\newblock \doi{10.1109/TNET.2017.2684909}

\bibitem{Cisco20}
Cisco: Cisco annual internet report (2020)

\bibitem{ir4}
Sabella, R., Iovanna, P., Bottari, G., Cavaliere, F.: Optical transport for
  industry 4.0.
\newblock J. Opt. Commun. Netw. \textbf{12}(8), 264--276 (2020).
\newblock \doi{10.1364/JOCN.390701}.
\newblock \urlprefix\url{http://jocn.osa.org/abstract.cfm?URI=jocn-12-8-264}

\bibitem{20years}
Winzer, P.J., Neilson, D.T., Chraplyvy, A.R.: Fiber-optic transmission and
  networking: the previous 20 and the next 20 years.
\newblock Opt. Express \textbf{26}(18), 24,190--24,239 (2018).
\newblock \doi{10.1364/OE.26.024190}.
\newblock
  \urlprefix\url{http://www.opticsexpress.org/abstract.cfm?URI=oe-26-18-24190}

\bibitem{new1}
Barla, I.B., Rambach, F., Schupke, D.A., Carle, G.: Efficient protection in
  single-domain networks using network coding.
\newblock In: 2010 IEEE Global Telecommunications Conference GLOBECOM 2010, pp.
  1--5 (2010).
\newblock \doi{10.1109/GLOCOM.2010.5683258}

\bibitem{hai_iet}
{Hai}, D.T., {Morvan}, M., {Gravey}, P.: Combining heuristic and exact
  approaches for solving the routing and spectrum assignment problem.
\newblock IET Optoelectronics \textbf{12}(2), 65--72 (2018).
\newblock \doi{10.1049/iet-opt.2017.0013}

\bibitem{hai_wiley}
Dao, H., Morvan, M., Gravey, P.: An efficient network-side path protection
  scheme in ofdm-based elastic optical networks.
\newblock International Journal of Communication Systems \textbf{31}(1), e3410.
\newblock \doi{10.1002/dac.3410}.
\newblock
  \urlprefix\url{https://onlinelibrary.wiley.com/doi/abs/10.1002/dac.3410}.
\newblock E3410 dac.3410

\bibitem{hai_nics}
Nguyen, D.M., Ngoc, L.A., Huong, P.T.V., Son, N.H., Hai, D.T.: An efficient
  column generation approach for solving the routing and spectrum assignment
  problem in elastic optical networks.
\newblock In: 2019 6th NAFOSTED Conference on Information and Computer Science
  (NICS), pp. 130--135 (2019).
\newblock \doi{10.1109/NICS48868.2019.9023831}

\bibitem{EON}
Lord, A., Zhou, Y.R., Jensen, R., Morea, A., Ruiz, M.: Evolution from
  Wavelength-Switched to Flex-Grid Optical Networks, pp. 7--30.
\newblock Springer International Publishing, Cham (2016)

\bibitem{new4}
Agrell, E., Karlsson, M., Chraplyvy, A.R., Richardson, D.J., Krummrich, P.M.,
  Winzer, P., Roberts, K., Fischer, J.K., Savory, S.J., Eggleton, B.J.,
  Secondini, M., Kschischang, F.R., Lord, A., Prat, J., Tomkos, I., Bowers,
  J.E., Srinivasan, S., Brandt-Pearce, M., Gisin, N.: Roadmap of optical
  communications.
\newblock Journal of Optics \textbf{18}(6), 063,002 (2016).
\newblock \doi{10.1088/2040-8978/18/6/063002}.
\newblock \urlprefix\url{https://doi.org/10.1088/2040-8978/18/6/063002}

\bibitem{new2}
Pasic, A., Tapolcai, J., Babarczi, P., Bérczi-Kovács, E.R., Király, Z.,
  Rónyai, L.: Survivable routing meets diversity coding.
\newblock In: 2015 IFIP Networking Conference (IFIP Networking), pp. 1--9
  (2015).
\newblock \doi{10.1109/IFIPNetworking.2015.7145330}

\bibitem{sla1}
Agrawal, A., Vyas, U., Bhatia, V., Prakash, S.: Sla-aware differentiated qos in
  elastic optical networks.
\newblock Optical Fiber Technology \textbf{36}, 41--50 (2017).
\newblock \doi{https://doi.org/10.1016/j.yofte.2017.01.012}

\bibitem{qos1}
Hai, D.T., Minh, H.T., Chau, L.H.: Qos-aware protection in elastic optical
  networks with distance-adaptive and reconfigurable modulation formats.
\newblock Optical Fiber Technology \textbf{61}, 102,364 (2021).
\newblock \doi{https://doi.org/10.1016/j.yofte.2020.102364}

\bibitem{hai_comletter}
Hai, D.T.: Leveraging the survivable all-optical wdm network design with
  network coding assignment.
\newblock IEEE Communications Letters \textbf{21}(10), 2190--2193 (2017).
\newblock \doi{10.1109/LCOMM.2017.2720661}

\bibitem{hai_oft}
Dao, T.H.: On optimal designs of transparent wdm networks with 1+1 protection
  leveraged by all-optical xor network coding schemes.
\newblock Optical Fiber Technology \textbf{40}, 93 -- 100 (2018).
\newblock \doi{https://doi.org/10.1016/j.yofte.2017.11.009}

\bibitem{hai_access}
Hai, D.T.: An optimal design framework for 1+1 routing and network coding
  assignment problem in wdm optical networks.
\newblock IEEE Access \textbf{5}, 22,291--22,298 (2017).
\newblock \doi{10.1109/ACCESS.2017.2761809}

\bibitem{hai_comcom}
Hai, D.T.: A bi-objective integer linear programming model for the routing and
  network coding assignment problem in wdm optical networks with dedicated
  protection.
\newblock Computer Communications \textbf{133}, 51 -- 58 (2019).
\newblock \doi{https://doi.org/10.1016/j.comcom.2018.08.006}

\bibitem{hai_comcom2}
Hai, D.T.: On routing, spectrum and network coding assignment problem for
  transparent flex-grid optical networks with dedicated protection.
\newblock Computer Communications  (2019).
\newblock \doi{https://doi.org/10.1016/j.comcom.2019.08.005}

\bibitem{hai_systems}
Hai, D.T., Chau, L.H., Hung, N.T.: A priority-based multiobjective design for
  routing, spectrum, and network coding assignment problem in
  network-coding-enabled elastic optical networks.
\newblock IEEE Systems Journal \textbf{14}(2), 2358--2369 (2020).
\newblock \doi{10.1109/JSYST.2019.2938590}

\bibitem{nc_others1}
Mitsolidou, C., Pleros, N., Miliou, A.: Digital all-optical physical-layer
  network coding for 2gbaud dqpsk signals in mm-wave radio-over-fiber networks.
\newblock Optical Switching and Networking \textbf{33}, 199--207 (2019).
\newblock \doi{https://doi.org/10.1016/j.osn.2017.10.002}

\bibitem{nc_others2}
Guan, X., Yang, Q., Wang, T., Chan, C.C.K.: Phase-aligned physical-layer
  network coding in visible light communications.
\newblock IEEE Photonics Journal \textbf{11}(2), 1--9 (2019).
\newblock \doi{10.1109/JPHOT.2019.2904954}

\bibitem{nc_others3}
Yanmei, J., Congmin, L., Pengfei, S., Lu, L.: Modulated retro-reflector-based
  physical-layer network coding for space optical communications.
\newblock IEEE Access \textbf{9}, 44,868--44,880 (2021).
\newblock \doi{10.1109/ACCESS.2021.3067101}

\bibitem{NC}
Ahlswede, R., et~al.: Network information flow.
\newblock Information Theory, IEEE Transactions on \textbf{46}(4), 1204--1216
  (2000).
\newblock \doi{10.1109/18.850663}

\bibitem{hai_rtuwo}
{Hai}, D.T.: Re-designing dedicated protection in transparent wdm optical
  networks with xor network coding.
\newblock In: 2018 Advances in Wireless and Optical Communications (RTUWO), pp.
  118--123 (2018).
\newblock \doi{10.1109/RTUWO.2018.8587873}

\bibitem{hai_springer}
Hai, D.T.: On solving the 1 + 1 routing, wavelength and network coding
  assignment problem with a bi-objective integer linear programming model.
\newblock Telecommunication Systems \textbf{71}(2), 155--165 (2019).
\newblock \doi{10.1007/s11235-018-0474-9}.
\newblock \urlprefix\url{https://doi.org/10.1007/s11235-018-0474-9}

\bibitem{hai_springer2}
Hai, D.T.: Network coding for improving throughput in wdm optical networks with
  dedicated protection.
\newblock Optical and Quantum Electronics \textbf{51}(387) (2019).
\newblock \doi{10.1007/s11082-019-2104-5}.
\newblock \urlprefix\url{https://doi.org/10.1007/s11082-019-2104-5}

\bibitem{sla2}
Layec, P., Dupas, A., Bisson, A., Bigo, S.: Qos-aware protection in flexgrid
  optical networks.
\newblock J. Opt. Commun. Netw. \textbf{10}(1), A43--A50 (2018).
\newblock \doi{10.1364/JOCN.10.000A43}.
\newblock \urlprefix\url{http://jocn.osa.org/abstract.cfm?URI=jocn-10-1-A43}

\bibitem{qos2}
Hai, D.T.: On the spectrum-efficiency of qos-aware protection in elastic
  optical networks.
\newblock Optik \textbf{202}, 163,563 (2020).
\newblock \doi{https://doi.org/10.1016/j.ijleo.2019.163563}

\bibitem{s7}
Zhu, F., Zhang, C., Zheng, Z., Farouk, A.: Practical network coding
  technologies and softwarization in wireless networks.
\newblock IEEE Internet of Things Journal \textbf{8}(7), 5211--5218 (2021).
\newblock \doi{10.1109/JIOT.2021.3056580}

\bibitem{s1}
Belzner, M., Haunstein, H., van Wijngaarden, A.J.: On the performance of
  network coding with protection cycles.
\newblock In: 2011 IEEE International Conference on Communications (ICC), pp.
  1--6 (2011).
\newblock \doi{10.1109/icc.2011.5963305}

\bibitem{s2}
Aly, S.A., Kamal, A.E.: Network coding-based protection strategies against a
  single link failure in optical networks.
\newblock In: 2008 International Conference on Computer Engineering Systems,
  pp. 251--256 (2008).
\newblock \doi{10.1109/ICCES.2008.4773006}

\bibitem{s5}
Zhijian, Q., Xianwei, Z., Shaojian, S., Yanfeng, C., Mingbo, Z.: Network coding
  based all-optical multicast in wdm networks.
\newblock The Journal of China Universities of Posts and Telecommunications
  \textbf{22}(1), 89--94 (2015).
\newblock \doi{https://doi.org/10.1016/S1005-8885(15)60630-6}.
\newblock
  \urlprefix\url{https://www.sciencedirect.com/science/article/pii/S1005888515606306}

\bibitem{s6}
Yang, L., Gong, L., Zhou, F., Cousin, B., Molnár, M., Zhu, Z.: Leveraging
  light forest with rateless network coding to design efficient all-optical
  multicast schemes for elastic optical networks.
\newblock Journal of Lightwave Technology \textbf{33}(18), 3945--3955 (2015).
\newblock \doi{10.1109/JLT.2015.2457092}

\bibitem{s3}
Beldachi, A.F., Anastasopoulos, M., Manolopoulos, A., Tzanakaki, A., Nejabati,
  R., Simeondou, D.: Resilient cloud-rans adopting network coding.
\newblock In: A.~Tzanakaki, M.~Varvarigos, R.~Mu{\~{n}}oz, R.~Nejabati,
  N.~Yoshikane, M.~Anastasopoulos, J.~Marquez-Barja (eds.) Optical Network
  Design and Modeling, pp. 349--361. Springer International Publishing, Cham
  (2020)

\bibitem{s4}
Savva, G., Manousakis, K., Ellinas, G.: Confidentiality meets protection in
  elastic optical networks.
\newblock Optical Switching and Networking \textbf{42}, 100,620 (2021).
\newblock \doi{https://doi.org/10.1016/j.osn.2021.100620}.
\newblock
  \urlprefix\url{https://www.sciencedirect.com/science/article/pii/S1573427721000175}

\bibitem{nc_others4}
Feng, N., Sun, X.: Implementation of network-coding approach for improving the
  ber performance in non-orthogonal multiple access (noma)-pon.
\newblock Optics Communications \textbf{462}, 125,301 (2020).
\newblock \doi{https://doi.org/10.1016/j.optcom.2020.125301}

\bibitem{nc_others5}
Lin, R., Cheng, Y., Guan, X., Tang, M., Liu, D., Chan, C.K., Chen, J.:
  Physical-layer network coding for passive optical interconnect in datacenter
  networks.
\newblock Opt. Express \textbf{25}(15), 17,788--17,797 (2017).
\newblock \doi{10.1364/OE.25.017788}.
\newblock
  \urlprefix\url{http://www.opticsexpress.org/abstract.cfm?URI=oe-25-15-17788}

\bibitem{nc_others6}
Engelmann, A., Bziuk, W., Jukan, A., Médard, M.: Exploiting parallelism with
  random linear network coding in high-speed ethernet systems.
\newblock IEEE/ACM Transactions on Networking \textbf{26}(6), 2829--2842
  (2018).
\newblock \doi{10.1109/TNET.2018.2852562}

\bibitem{nc_others8}
El~Asghar, N.B., Jouili, I., Frikha, M.: Survivable inter-datacenter network
  design based on network coding.
\newblock In: 2017 IEEE/ACS 14th International Conference on Computer Systems
  and Applications (AICCSA), pp. 1192--1197 (2017).
\newblock \doi{10.1109/AICCSA.2017.113}

\bibitem{sla3}
Layec, P., Dupas, A., Bisson, A., Bigo, S.: Qos-aware protection in flexgrid
  optical networks.
\newblock In: 2017 Optical Fiber Communications Conference and Exhibition
  (OFC), pp. 1--3 (2017)

\bibitem{pp1}
Sivakumar, M., Sivalingam, K.M., Somani, A.K.: Partial protection in optical
  wdm networks: Enhanced support for dynamic traffic.
\newblock In: 2006 3rd International Conference on Broadband Communications,
  Networks and Systems, pp. 1--8 (2006).
\newblock \doi{10.1109/BROADNETS.2006.4374328}

\bibitem{pp2}
Kuperman, G., Modiano, E., Narula-Tam, A.: Partial protection in networks with
  backup capacity sharing.
\newblock In: National Fiber Optic Engineers Conference, p. NW3K.4. Optical
  Society of America (2012).
\newblock \doi{10.1364/NFOEC.2012.NW3K.4}.
\newblock
  \urlprefix\url{http://www.osapublishing.org/abstract.cfm?URI=NFOEC-2012-NW3K.4}

\bibitem{pp3}
Savas, S.S., Ma, C., Tornatore, M., Mukherjee, B.: Backup reprovisioning with
  partial protection for disaster-survivable software-defined optical networks
  \textbf{31}(2) (2016).
\newblock \doi{10.1007/s11107-015-0563-6}.
\newblock \urlprefix\url{https://doi.org/10.1007/s11107-015-0563-6}

\bibitem{pp5}
Zhong, Z., Li, J., Hua, N., Figueiredo, G.B., Li, Y., Zheng, X., Mukherjee, B.:
  On qos-assured degraded provisioning in service-differentiated multi-layer
  elastic optical networks.
\newblock In: 2016 IEEE Global Communications Conference (GLOBECOM), pp. 1--5
  (2016).
\newblock \doi{10.1109/GLOCOM.2016.7842043}

\bibitem{pp6}
Nassar, B.O., Tachibana, T.: Degraded provisioning of spectrum and holding time
  with qos assurance in elastic optical networks.
\newblock In: 2019 24th OptoElectronics and Communications Conference (OECC)
  and 2019 International Conference on Photonics in Switching and Computing
  (PSC), pp. 1--3 (2019).
\newblock \doi{10.23919/PS.2019.8817760}

\bibitem{pp7}
Lisboa, F., Fonseca, K.V.O., Vieira, L.C., Monti, P., Figueiredo, G.B.,
  de~Santi, J.: Restoration based on bandwidth degradation and service
  restoration delay for optical cloud networks.
\newblock In: GLOBECOM 2017 - 2017 IEEE Global Communications Conference, pp.
  1--6 (2017).
\newblock \doi{10.1109/GLOCOM.2017.8254773}

\bibitem{pp8}
Vela, A.P., Ruiz, M., Cugini, F., Velasco, L.: Combining a machine learning and
  optimization for early pre-fec ber degradation to meet committed qos.
\newblock In: 2017 19th International Conference on Transparent Optical
  Networks (ICTON), pp. 1--4 (2017).
\newblock \doi{10.1109/ICTON.2017.8025009}

\end{thebibliography}

\end{document}